\documentclass[12pt]{article}
\usepackage{a4wide}
\usepackage{epsfig}

\def\be{\begin{equation}}
\def\ee{\end{equation}}
\def\ba{\begin{eqnarray}}
\def\ea{\end{eqnarray}}

\def\lsim{\raise0.3ex\hbox{$\;<$\kern-0.75em\raise-1.1ex\hbox{$\sim\;$}}}
\def\gsim{\raise0.3ex\hbox{$\;>$\kern-0.75em\raise-1.1ex\hbox{$\sim\;$}}}

\def\nuebar{{\bar{\nu}_e}}
\def\nux{{\nu_x}}

\def\ebar{{\bar{e}}}
\def\xbar{{\bar{x}}}

\def\la{\langle}
\def\ra{\rangle}

\def\theta{\vartheta}

\begin{document}

\quad\vskip1cm
\begin{center}
{\bf\large 
Identifying the neutrino mass hierarchy\\ \vskip0.1cm
with supernova neutrinos}
\vskip0.5cm
{M.~Kachelrie\ss\ and R.~Tom\`as}
\\
{\small\it Max-Planck-Institut f\"ur Physik (Werner-Heisenberg-Institut),
M\"unchen}
\end{center}

\begin{abstract}
We review how a high-statistics observation of the neutrino signal
from a future galactic core-collapse supernova (SN) may be used to
discriminate between different 
neutrino mixing scenarios. Most SN neutrinos are emitted in the
accretion and cooling phase, during which the flavor-dependent
differences of the emitted neutrino spectra are small and rather
uncertain. Therefore the discrimination between neutrino mixing
scenarios using these neutrinos
should rely  on observables independent of the
SN neutrino spectra. We discuss two complementary methods that allow
for the positive identification of the mass hierarchy without knowledge of 
the emitted neutrino fluxes, provided that the 13-mixing angle is
large, $\sin^2\theta_{13}\gg 10^{-5}$. 
These two approaches are the observation of modulations in the
neutrino spectra by Earth matter effects or by the passage of
shock waves through the SN envelope.
If the value of the 13-mixing angle is unknown, using additionally the
information encoded in the prompt neutronization $\nu_e$
burst---a robust feature found in all modern SN simulations---can be
sufficient to fix both the neutrino hierarchy and to 
decide whether $\theta_{13}$ is ``small'' or ``large.''     
\end{abstract}

%%%%%%%%%%%%%%%%%%%%%%%%%%%%%%%%%%%%%%%%%%%%%%%%%%%%%%%%%%%%%%%%%%%%
\section{Introduction}
\label{intro}
%%%%%%%%%%%%%%%%%%%%%%%%%%%%%%%%%%%%%%%%%%%%%%%%%%%%%%%%%%%%%%%%%%%%

Despite the enormous progress of neutrino physics in the last decade, 
many open questions remain to be solved. Among them are two, the 
mass hierarchy---normal versus inverted mass spectrum---and the
value of the 13-mixing angle $\theta_{13}$, where the observation of
neutrinos from a  core-collapse supernova (SN) could provide
important clues~\cite{Dighe:1999bi,Lunardini:2003eh,Takahashi:2003bj}. 
Schematically, the neutrino emission by a SN can be divided
into four stages: Infall phase, neutronization burst, accretion, 
and Kelvin-Helmholtz cooling phase. Most SN neutrinos are emitted
during the last two stages,  
 in all flavors with only small
differences between the $\bar\nu_e$ and $\bar\nu_{\mu,\tau}$ 
spectra. Moreover, the absolute values of the average neutrino
energies as well as the relative size of the luminosities during the
accretion and cooling phases cannot be determined with sufficient precision,
especially if the SN is optically obscured and the progenitor type
remains unknown.
Therefore, a straightforward extraction of oscillation parameters
from the bulk of the SN neutrino signal 
seems hopeless. Only features in the
detected neutrino spectra that are independent of unknown SN
parameters should be used in such an analysis.

The two most promising sources for such features are the  
modulations in the neutrino spectra caused by the Earth matter or by
the passage of shock waves through the SN envelope. In the first case, 
matter effects on SN neutrinos traversing the Earth give rise to
specific frequencies in the energy spectrum of these
neutrinos, which are analytically known and depend only on the
neutrino properties and the distance traveled through the
Earth~\cite{ceciliaearth,Dighe:2003jg,Dighe:2003vm}. 
In the other case, the passage of the SN shock waves through the density 
region corresponding to resonant neutrino oscillations with the atmospheric
neutrino mass difference imprints specific time- and
energy-dependent modulations on the neutrino energy 
spectrum~\cite{Schirato,Tomas:2004gr}, difficult to be mimicked
by other effects. 
Only the amplitude of both modulations, and thus the statistical
confidence to detect them, depends on how different the emitted
neutrino fluxes are, while the specific shape of the modulations 
is independent from the fluxes.

\begin{table}[ht]
\begin{center}
\begin{tabular}{llcccc}
\hline
Case & Hierarchy &  $\sin^2 \theta_{13}$ & Earth & Shock & $\nu_e$ burst  \\
\hline
A &  Normal & $\gsim 10^{-3}$ & Yes & No & No \\
B & Inverted &  $\gsim 10^{-3}$ & No & Yes & Yes  \\
C & Any & $\lsim 10^{-5}$ & Yes & No & Yes \\
\hline
\end{tabular}
\end{center}
\caption{The presence of Earth-matter and shock wave effects in the
  $\bar\nu_e$ spectra and of the $\nu_e$ burst for
  different neutrino mixing scenarios. \label{abc-table}}
\end{table}

\begin{figure}[h!]
\begin{center}

\epsfig{file=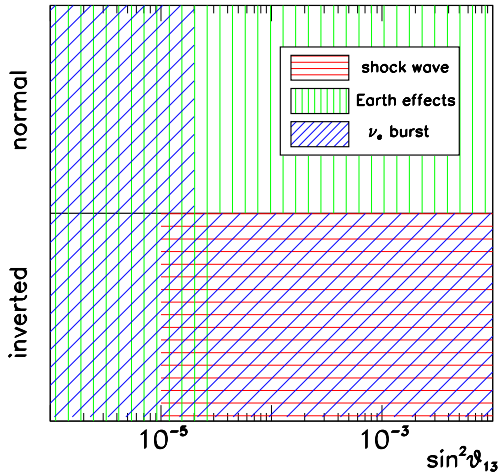,width=7cm}
\label{fig:box}
\caption{Sensitivity of the three SN observables discussed in the text
  to  different neutrino mixing scenarios.
}
\end{center}
\end{figure}

In the following sections we will concentrate on three different
neutrino mixing schemes (A, B, C), cf. Tab.~\ref{abc-table}, where
modulations by Earth or SN shock effects are clearly separated. 
For an inverted hierarchy and intermediate values of the 13-mixing angle, 
$10^{-5}\lsim \sin^2 \theta_{13} \lsim 10^{-3}$, 
both effects can be present. In this case, the value of $\theta_{13}$
has to be close to $10^{-5}$, the lower limit of the intermediate
range.
In Fig.~\ref{fig:box} we show schematically the
sensitivity of several SN observables to different neutrino mixing
parameters.

If at the time of the SN detection the value of $\theta_{13}$ is known
to be ``large,'' then the neutrino mass hierarchy can be identified
observing the modulations induced either by the SN shock wave
propagation (case B in Tab.~\ref{abc-table}) or by the Earth matter effects
(case A). 
If the value of $\theta_{13}$ is still unknown and Earth
matter effects are observed, an ambiguity exists between case A and
C. In sec.~4, we discuss why the neutrino emission during the
neutronization burst is only weakly dependent on variations in the
input of current SN models and how the neutronization burst can be
used to break the degeneracy between A and C.
 
%%%%%%%%%%%%%%%%%%%%%%%%%%%%%%%%%%%%%%%%%%%%%%%%%%%%%%%%%%%%%%%%%%%%%%%
\section{Identifying signatures of the SN shock wave propagation}

The neutrino spectra $F_{\nu_i}$ arriving at the Earth are 
determined by the primary neutrino spectra $F^0_{\nu_i}$ as well as
the neutrino mixing scenario, 
$F_{\nu_i}(E,t) = \sum_{j} p_{ji}(E,t) F^0_{\nu_j}(E,t)$,
where $p_{ji}$ is the conversion  probability of a $\nu_j$ into $\nu_i$
after propagation through the SN mantle.
The probabilities $p_{ji}$ are basically determined by the
number of resonances that the neutrinos traverse and their
adiabaticity. Both are directly connected to the neutrino mixing scheme.
In contrast to the solar case, SN neutrinos must pass through
two resonance layers: the H-resonance layer at 
$\rho_{\rm H}\sim 10^3$~g/cm$^3$ corresponding to $\Delta m^2_{\rm atm}$,
and the L-resonance layer at 
$\rho_{\rm L}\sim 10$~g/cm$^3$ corresponding to $\Delta m^2_{\odot}$.
Whereas the L-resonance is always adiabatic and 
in the neutrino channel, the adiabaticity of
the H-resonance depends on the value of $\theta_{13}$, and 
the resonance shows up in the neutrino or antineutrino channel 
for a normal or inverted mass hierarchy respectively~\cite{Dighe:1999bi}.

\begin{figure}[ht]
\epsfig{file=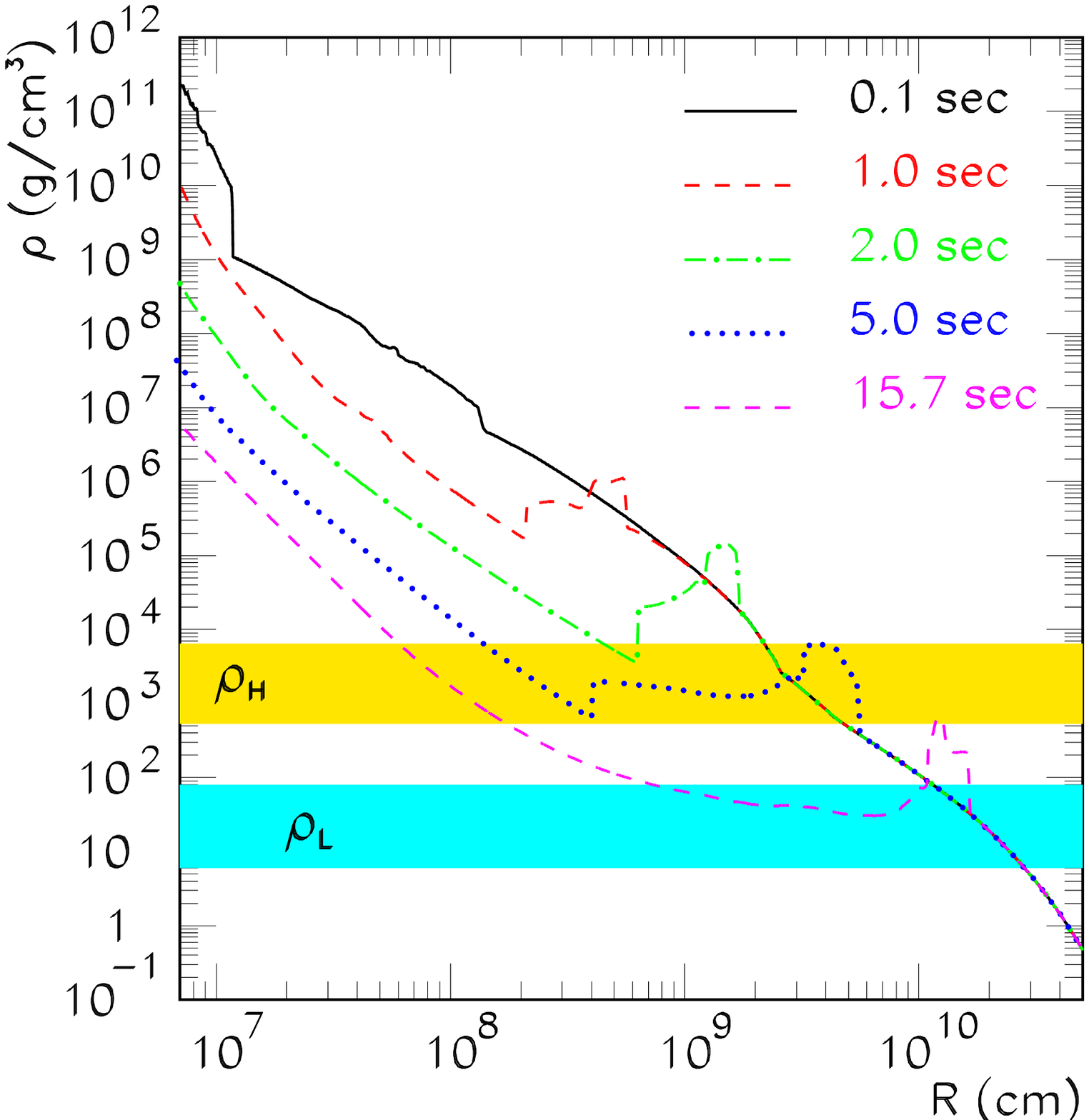,width=8cm}
\epsfig{file=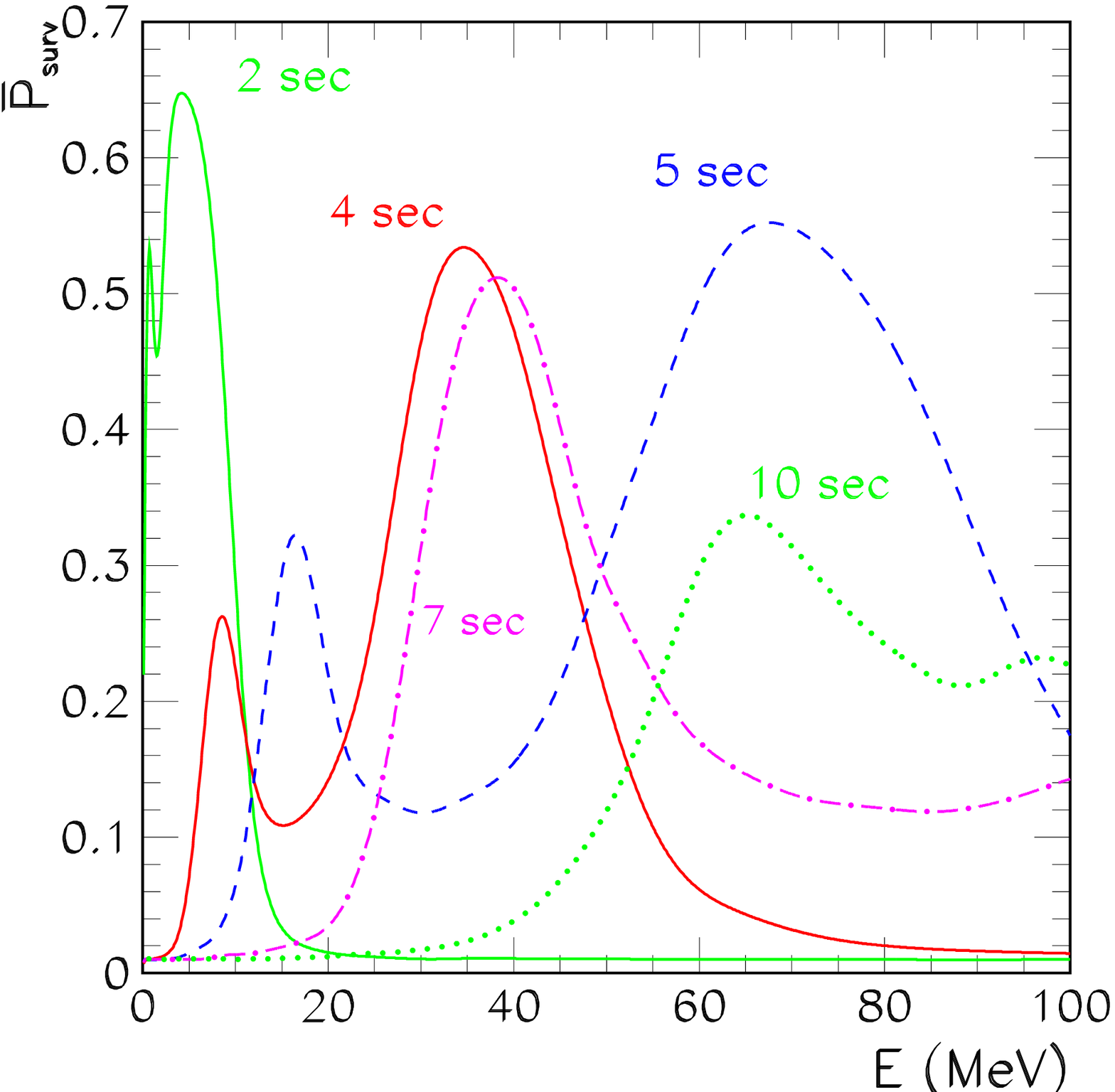,width=8cm}
\label{fig:shockpropagation}
\caption{
Left: 
Shock and reverse-shock propagation. The density profile is shown
at the indicated instances after core bounce. The density region
$\rho_{\rm H}$ corresponds to resonant neutrino oscillations with the
atmospheric mass difference, $\rho_{\rm L}$ to the solar
one~\cite{Tomas:2004gr}. 
Right: 
Survival probability $\bar p(E,t)$ as function of energy at
different times~\cite{Tomas:2004gr}.
\label{pee}
}
\end{figure}

During approximately the first two seconds after core bounce, the
neutrino survival probabilities are constant in time and in energy for
all three cases A, B, and C.
However, at $t\approx 2$ s the H-resonance layer is reached by the
outgoing shock wave, see the left panel of Fig.~\ref{pee}. 
The way the shock wave passage affects the neutrino propagation
strongly depends on the neutrino mixing scenario: 
cases A and C will not show any evidence of shock wave propagation
in the observed $\nuebar$ spectrum, either because there is no
resonance in the antineutrino channel as in scenario A, or because the
resonance is always strongly non-adiabatic as in scenario C.
However, in scenario B, the sudden change in the density breaks  the
adiabaticity of the resonance, leading to observable consequences in the
$\nuebar$ spectrum.

The key ingredient to observe signatures of the shock wave propagation is
the time and energy dependence of the neutrino survival probability.
In the right panel of Fig.~\ref{pee}, we show  $\bar p(E,t)\equiv
p_{\bar\nu_e\bar\nu_e} $ averaged
with the energy resolution function of Super-Kamiokande, for the case with
a forward and a reverse shock. The latter forms when a
neutrino-driven baryonic wind develops and collides with the 
earlier, more slowly expanding SN ejecta. 
Although the exact propagation history depends on the detailed
dynamics during the early stages of the SN explosion, 
a reverse shock forms in all models which
were computed with sufficient resolution~\cite{Tomas:2004gr}.  
The presence of two shocks results in a dip in $\bar p(E,t)$ at
those energies for which the resonance region is passed by both shock
waves. All these structures move in time towards higher energies, as
the shock waves reach regions with lower density.

\begin{figure}
\epsfig{file=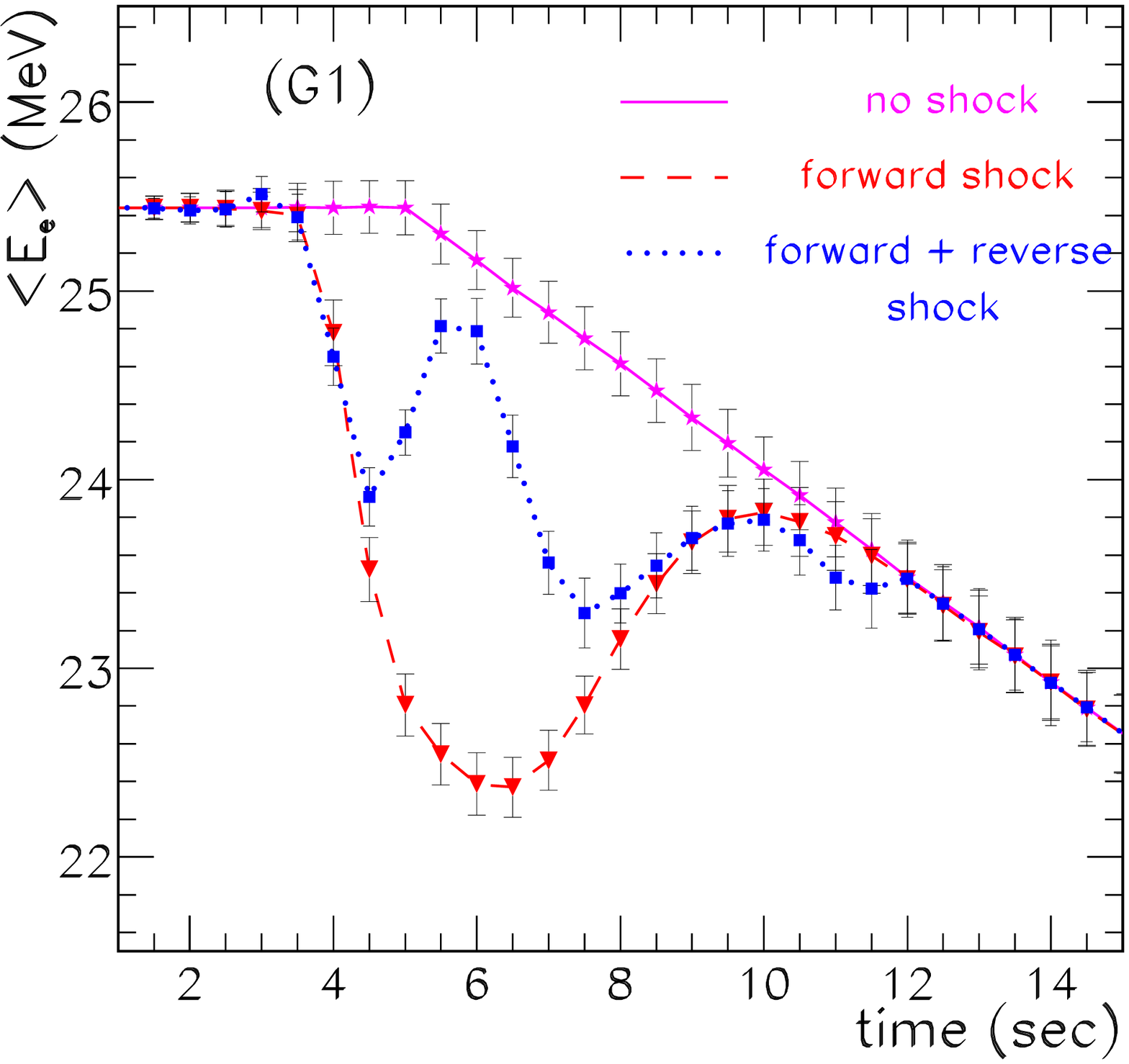,width=8cm,height=6.5cm}
\epsfig{file=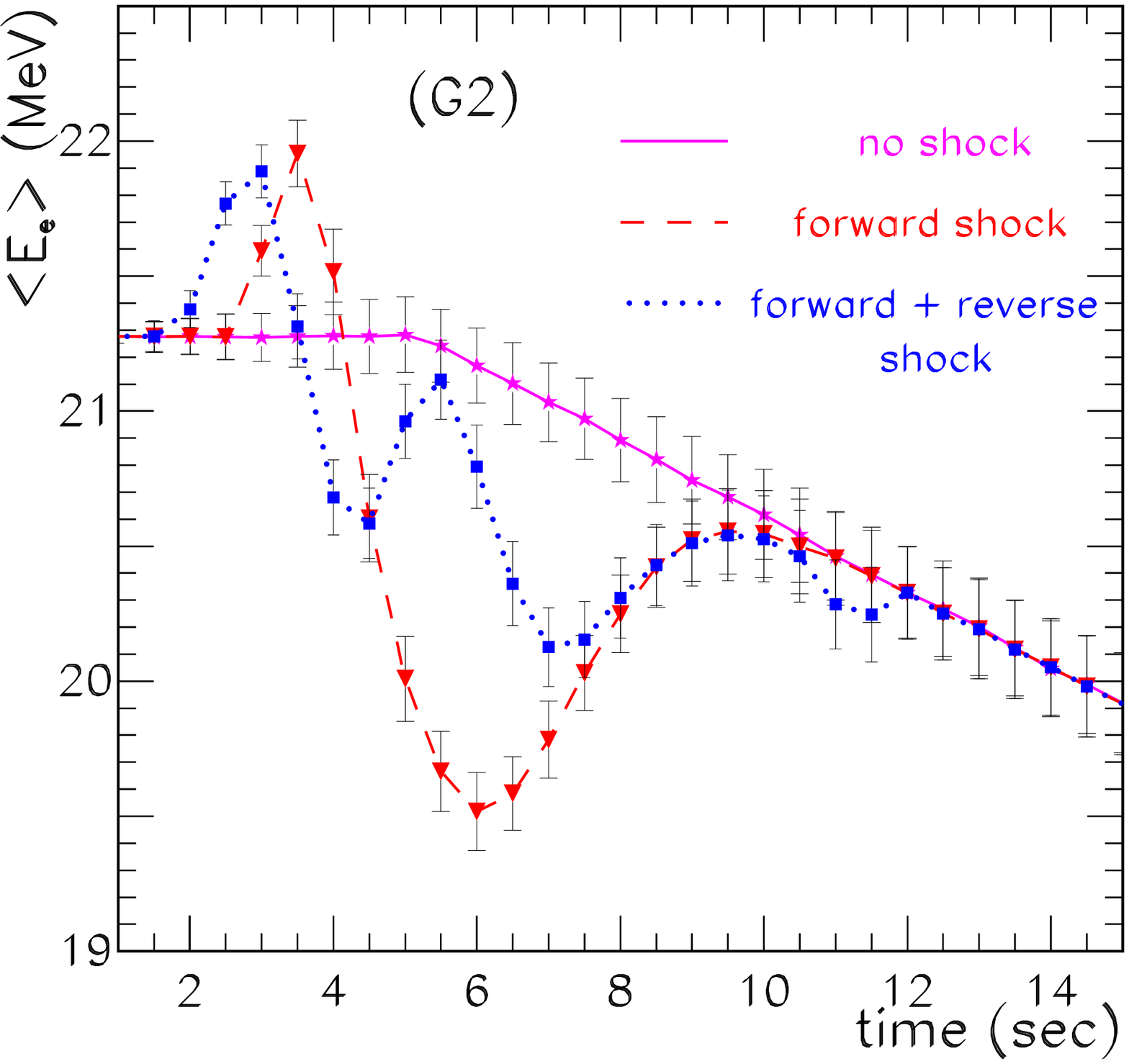,width=8cm,height=6.5cm}
\caption{The average energy of $\bar\nu p\to ne^+$ events binned in time 
for a static density profile (magenta), a profile with only a forward
shock (red) and with forward and reverse shock (blue). 
The error bars represent 1~$\sigma$ errors in any bin, from
Ref.~\cite{Tomas:2004gr}. 
\label{Eav}}
\end{figure}

A useful observable to detect effects of the shock propagation is the
average of the measured positron energies, $\la E_e\ra$, produced in
inverse beta decays $\bar\nu_e+p\to n+e^+$.
In Fig.~\ref{Eav}, we show $\la E_e \ra$
together with the one sigma errors expected for a megaton water
Cherenkov detector and a SN in 10~kpc distance, with a time binning of
0.5~s: Both panels contains the case that no shock wave
influences the neutrino propagation, the case of only a forward shock
wave and of both forward and reverse shock wave. 
The left and right panels show two different models for neutrino
fluxes: G1 assumes different average energies of the emitted neutrinos,
$\langle E_0(\nux) \rangle/\langle E_0(\nuebar) \rangle=1.2$, and
similar fluxes,  $\Phi_0(\nu_e)/\Phi_0(\nu_x)=0.8$, while G2
assumes identical energy spectra, $\langle E_0(\nux) \rangle/\langle
E_0(\nuebar) \rangle=1$, and $\Phi_0(\nu_e)/\Phi_0(\nu_x)=0.5$.

\begin{figure}
\epsfig{file=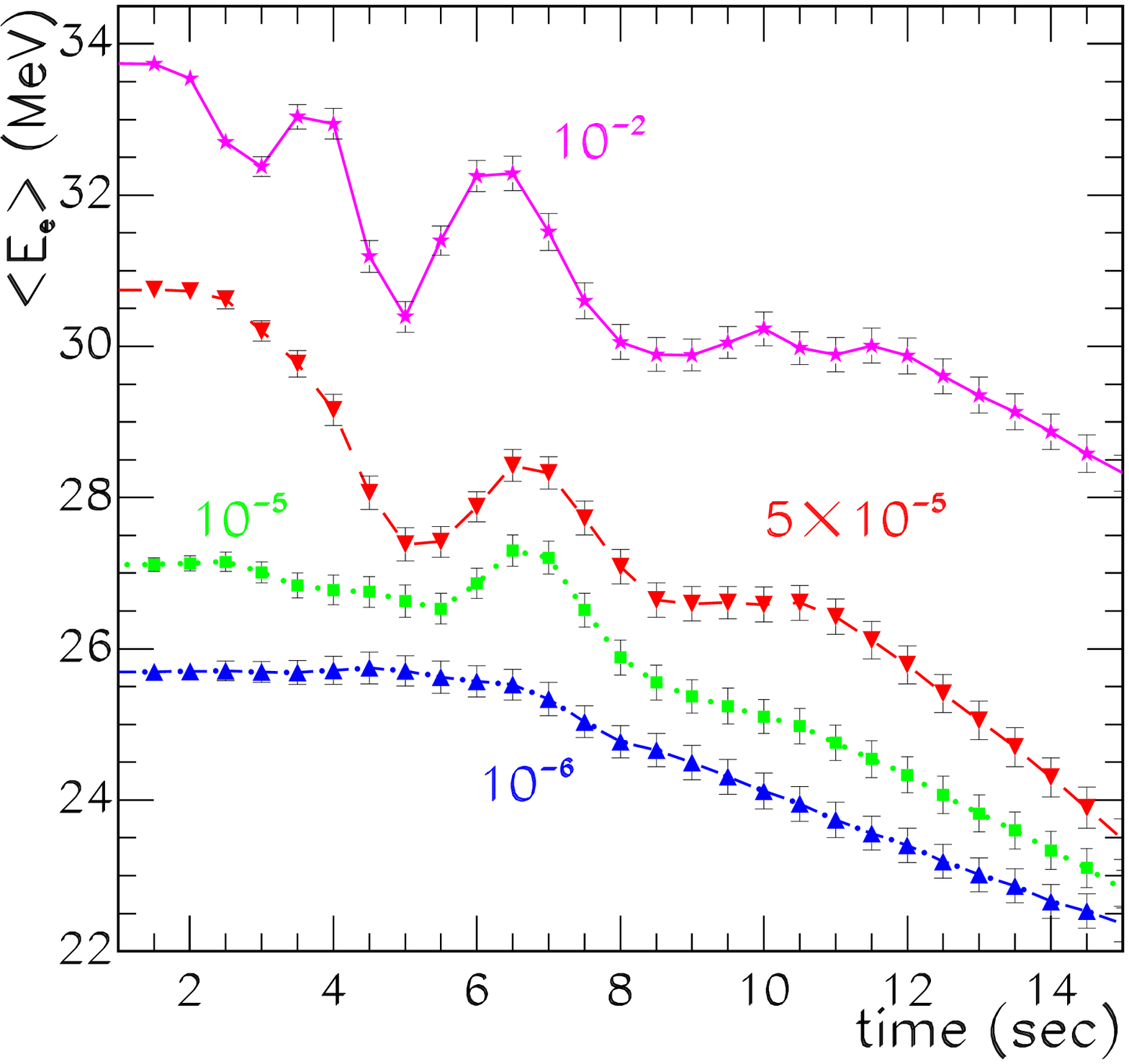,width=8cm,height=6.5cm}
\epsfig{file=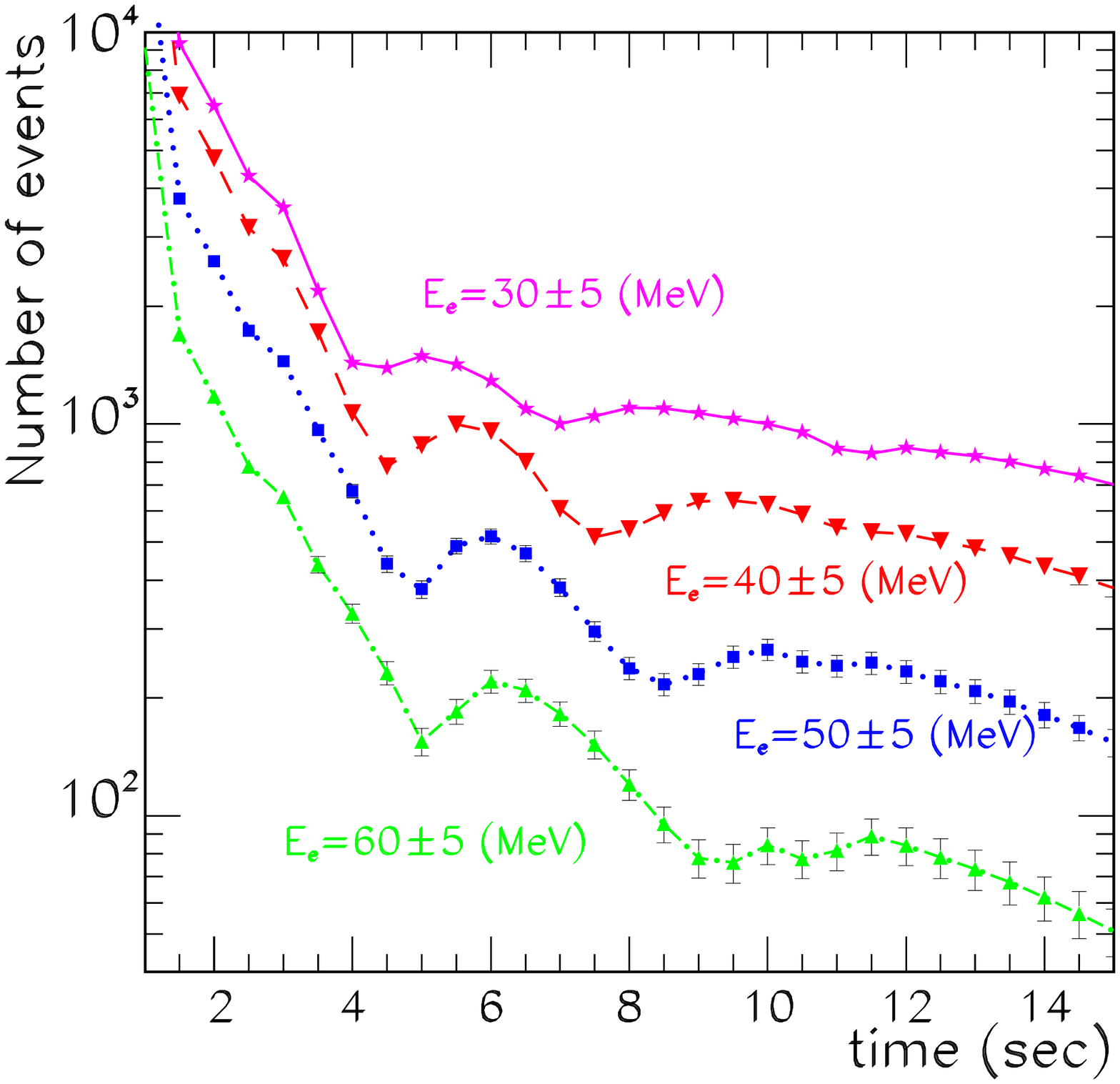,width=8cm,height=6.5cm}
\caption{
 Left: Time dependence of $\la E_e\ra$ 
  for a profile with a forward and reverse shock for
  several values of $\tan^2\theta_{13}$ as indicated.
  Right:  Number of events binned per energy decade as function
  of time for forward and reverse shock, from Ref.~\cite{Tomas:2004gr}.  
\label{alpha}}
\end{figure}

The effects of the shock wave propagation are clearly visible,
independent of the assumptions about the initial neutrino 
spectra. Moreover, it is not only possible to detect the shock wave
propagation in general, but also to identify the specific 
imprints of the forward and reverse shock versus the forward shock only  
case. The signature of the reverse shock is its double-dip structure
compared to the one-dip of a forward shock only.
To study the dependence of the double-dip structure on the value of
$\theta_{13}$, we show $\la E_e\ra$ as function of time for different
13-mixing angles in the left panel of Fig.~\ref{alpha}. Even for as
small values as $\tan^2\theta_{13}=5\times 10^{-5}$ the double-dip is
still clearly visible, while for $\tan^2\theta_{13}=1\times 10^{-5}$
only a bump modulates the neutrino signal.

In the right panel of Fig.~\ref{alpha}, we show the number of events binned
in energy intervals of 10~MeV as function of time for the case of a
reverse shock. We can observe clearly how the positions of the two dips
change in each energy bin. It is remarkable that the double-dip
feature allows one to trace 
the shock propagation: Given the neutrino mixing scheme, the neutrino
energy fixes the resonance density. Therefore, the progress of the
shock fronts can be read off from the position of the double-dip in
the neutrino spectra of different energy.
Thus, the observation of shock wave effects does not only identify
case B (inverted hierarchy, large $\theta_{13}$), but gives also
access to physics deep inside the SN.

%%%%%%%%%%%%%%%%%%%%%%%%%%%%%%%%%%%%%%%%%%%%%%%%%%%%%%%%%%%%%%%%%%%%%%%%%%
\section{Earth-matter effects}

During the first two seconds after post-bounce, during which
roughly half of all neutrinos are emitted, the dependence of the
probability to reach the Earth on the neutrino energy $E$ is very
weak. 
However, if neutrinos cross the Earth before reaching the detector, 
the conversion probabilities may become energy-dependent and induce
modulations in the neutrino energy spectrum. These
modulations  may be observed in the form of 
local peaks and valleys in the spectrum of the event rate $\sigma
F_\ebar^D$ plotted as a function of $1/E$. 
These modulations arise in the antineutrino channel only in cases A
and C. Therefore its observation would exclude case B.
This distortion in the spectra could be  measured by
comparing the 
neutrino signal at two or more different detectors such that the
neutrinos travel different distances through the Earth before reaching
them~\cite{ceciliaearth,Dighe:2003be}. 
However these Earth matter effects can be also identified in a single
detector~\cite{Dighe:2003jg,Dighe:2003vm}.

\begin{figure}
\epsfig{file=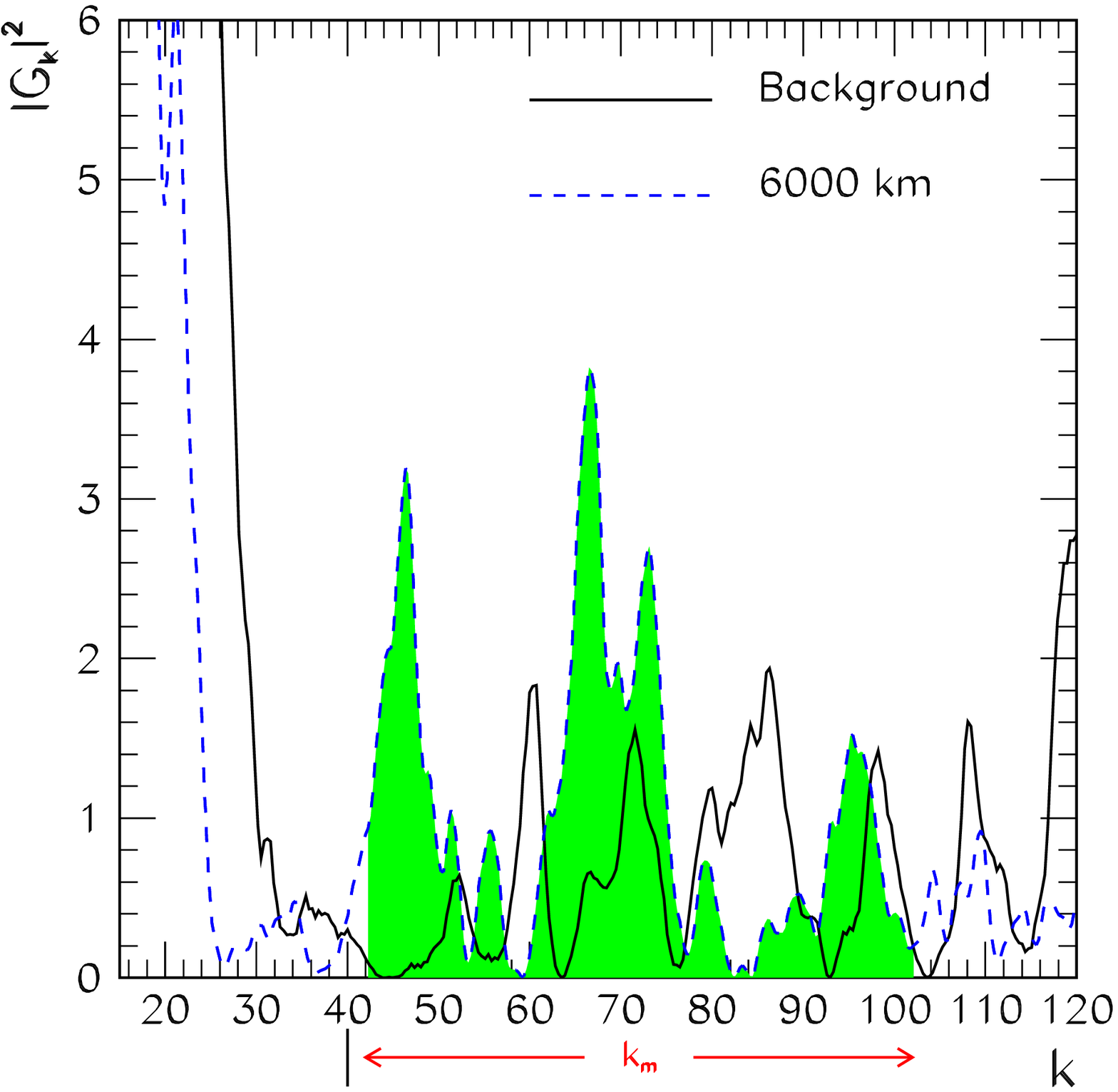,width=8cm,height=6.5cm}
\epsfig{file=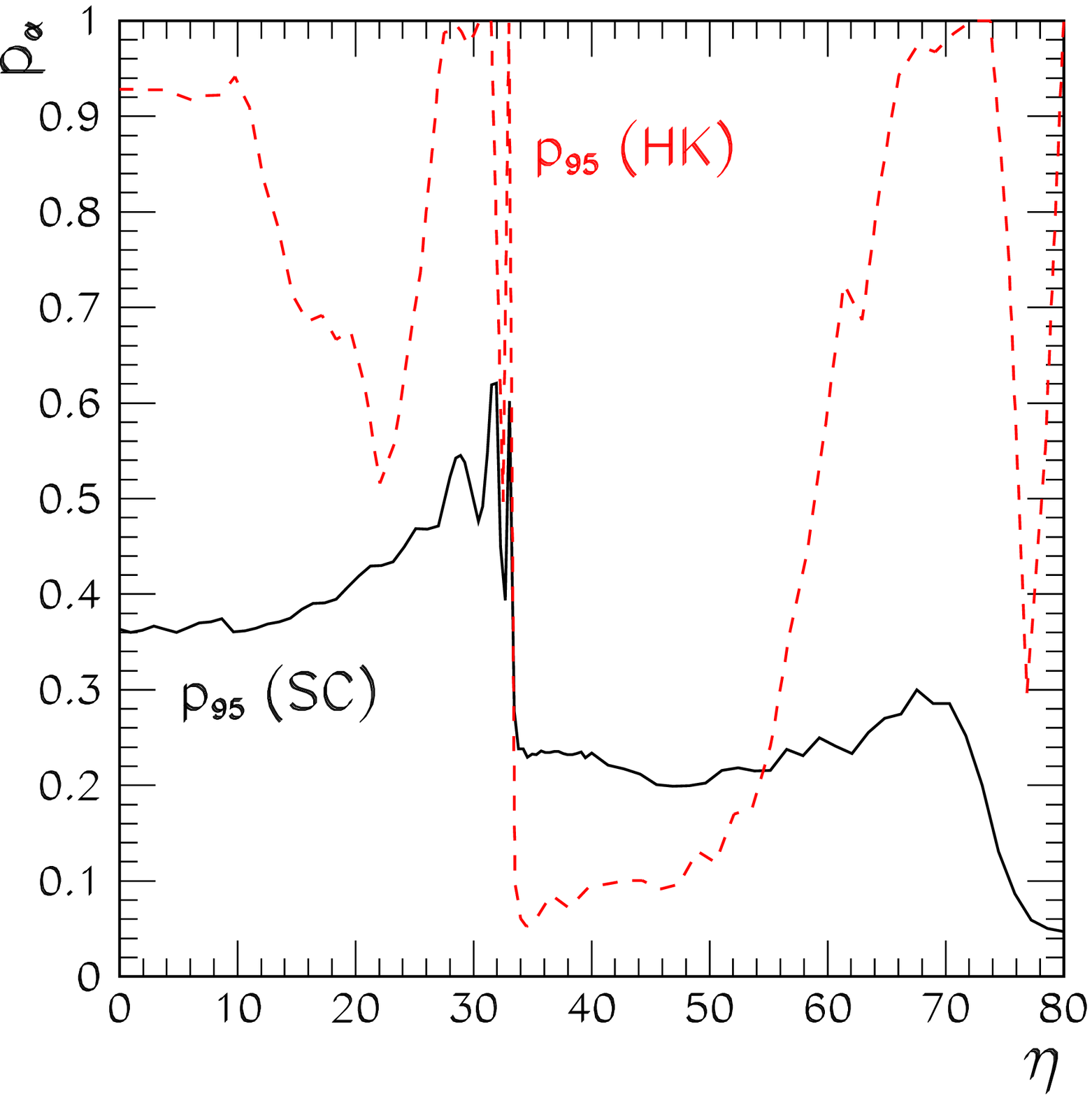,width=8cm,height=6.5cm}
\caption{
Left: Realistic power spectrum from a single simulation~\cite{Dighe:2003vm}.
Right: Comparison of $p_{95}$ as a function of nadir angle $\eta$ using
a ``floating cut'' as discussed in Ref.~\cite{Dighe:2003vm} 
for a 32 kton scintillator (SC) and a megaton
water Cherenkov (HK) detector.
\label{w}}
\end{figure}

The net $\nuebar$ flux at the detector may be written in the form
\be
F_{\bar{e}}^D =  \sin^2 \theta_{12}  F_\xbar^0 + 
\cos^2 \theta_{12} F_{\bar{e}}^0
+ \Delta F^0
\sum_{i=1}^7 \bar{A}_i \sin^2(k_i y /2) \,,
\label{feDbar-y}
\ee
where $y$ is the ``inverse energy'' parameter $y\equiv 12.5~{\rm MeV}/E$,  
$\Delta F^0 \equiv (F_{\bar{e}}^0 - F_\xbar^0)$ depends
only on the primary neutrino spectra, whereas the $\bar{A}_i$ 
depend only on the mixing parameters and are independent of the
primary spectra.

The last term in Eq.~(\ref{feDbar-y}) is the Earth oscillation term 
that contains up to seven analytically known frequencies $k_i$ in
$y$, the coefficients $\Delta F^0 \bar{A}_i$ being relatively slowly
varying functions of $y$.  The first two terms in Eq.~(\ref{feDbar-y})
are also slowly varying functions of $y$, and hence contain
frequencies in $y$ that are much smaller than the $k_i$.  
The frequencies $k_i$ are completely independent of the primary
neutrino spectra, and can be determined to a good accuracy from
the knowledge of the solar oscillation parameters, the Earth matter 
density, and the position of the SN in the sky~\cite{Dighe:2003vm}.
The latter can be determined with sufficient precision 
even if the SN is optically obscured using the pointing capability of
water Cherenkov neutrino detectors~\cite{pointing}.

The power spectrum of $N$ detected neutrino events is
\begin{equation}
G(k) \equiv \frac{1}{N} \left| \sum_{i=1}^N e^{iky_i} \right|^2 \,.
\label{ft-def}
\end{equation}
In the absence of Earth effect modulations, $G(k)$ has an average
value of one for $k \gsim 40$.  The region $k \lsim 40$ is dominated by the
``0-peak'', cf. Fig.~\ref{w}, which is a manifestation of the low
frequency terms in Eq.~(\ref{feDbar-y}).  Identifying Earth effects is
equivalent to observing excess power in $G(k)$ around the known
frequencies $k_i$.
The area under the power
spectrum between two fixed frequencies $k_{\rm min}$ and $k_{\rm max}$
is on an average $(k_{\rm max}-k_{\rm min})$. In the absence of Earth
effects, this area will have a distribution centered around this mean.
The Earth effect peaks tend to increase this area. 
The confidence level of peak identification, $p_\alpha$, may then be
defined as the fraction of the area of the background distribution
that is less than the actual area measured.

In the right panel of Fig.~\ref{w}, we assume the model
G1 for the neutrino fluxes and compare the results
obtained with a 32 kton 
scintillator detector and a megaton water Cherenkov detector.  In the
latter case, as neutrinos travel more and more distance in the mantle
the peak moves to higher $k$ values, and due to the high $k$
suppression, the efficiency of
peak identification decreases. When the neutrinos start
traversing the core, additional low $k$ peaks are generated and the
efficiency increases again.

The identification of Earth matter effects excludes case B, and is
thus complementary to the observation of shock wave effects.

%%%%%%%%%%%%%%%%%%%%%%%%%%%%%%%%%%%%%%%%%%%%%%%%%%%%%%%%%%%%%%%%%%%%%%%%%
\section{Neutronization $\nu_e$ burst}

If the value of $\theta_{13}$ is unknown, a degeneracy exists between
case A and C. Both scenarios predict the same $\bar\nu_e$ signature
in a water Cherenkov detector, and therefore the previous two
observables are not useful to disentangle them.
In this case, the additional information
encoded in the $\nu_e$ neutrinos emitted during the neutronization
burst can fix the range of $\theta_{13}$ as well as the neutrino
mass hierarchy.  

\begin{figure}
\epsfig{file=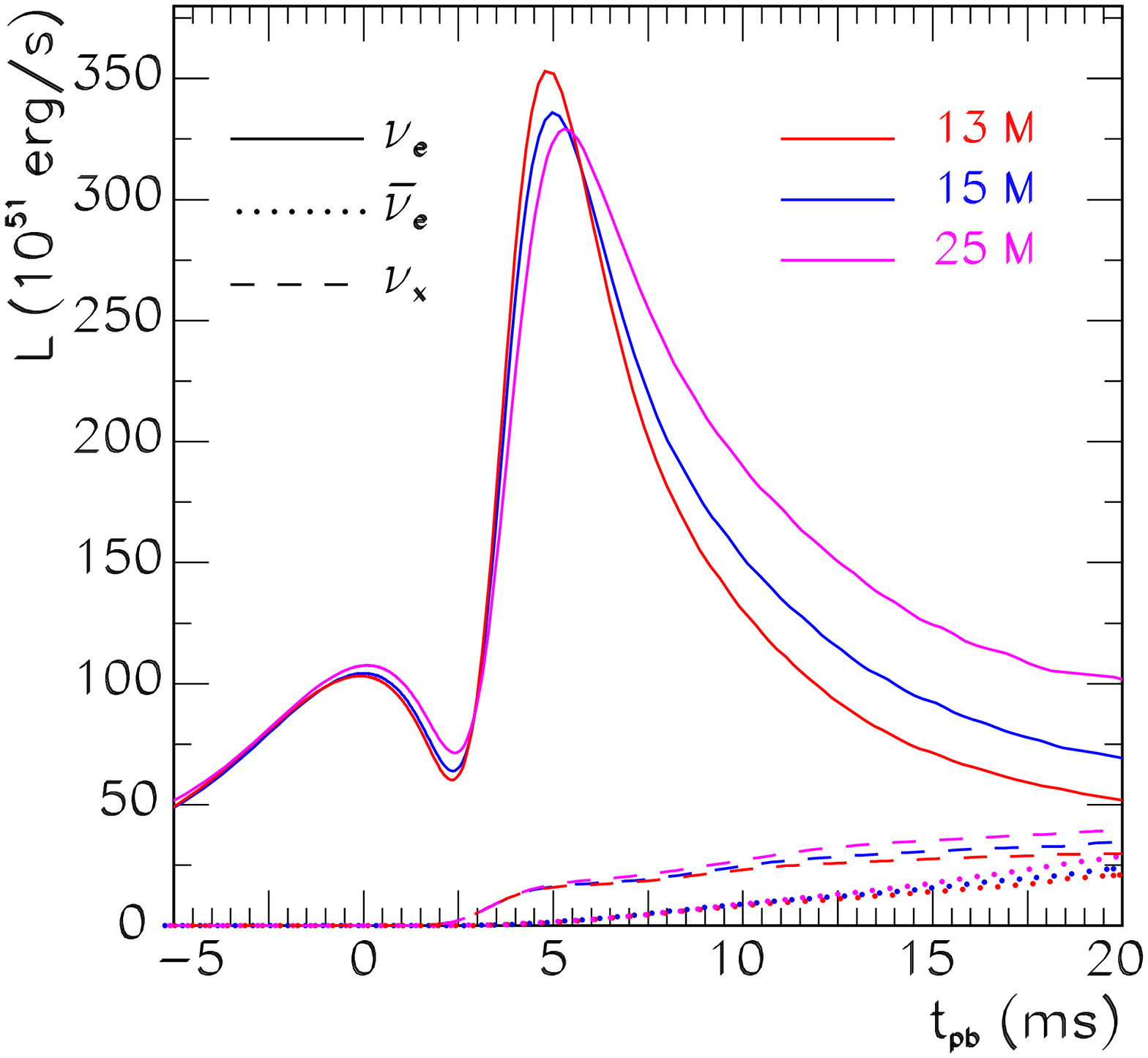,width=8.cm,height=6.5cm}
\epsfig{file=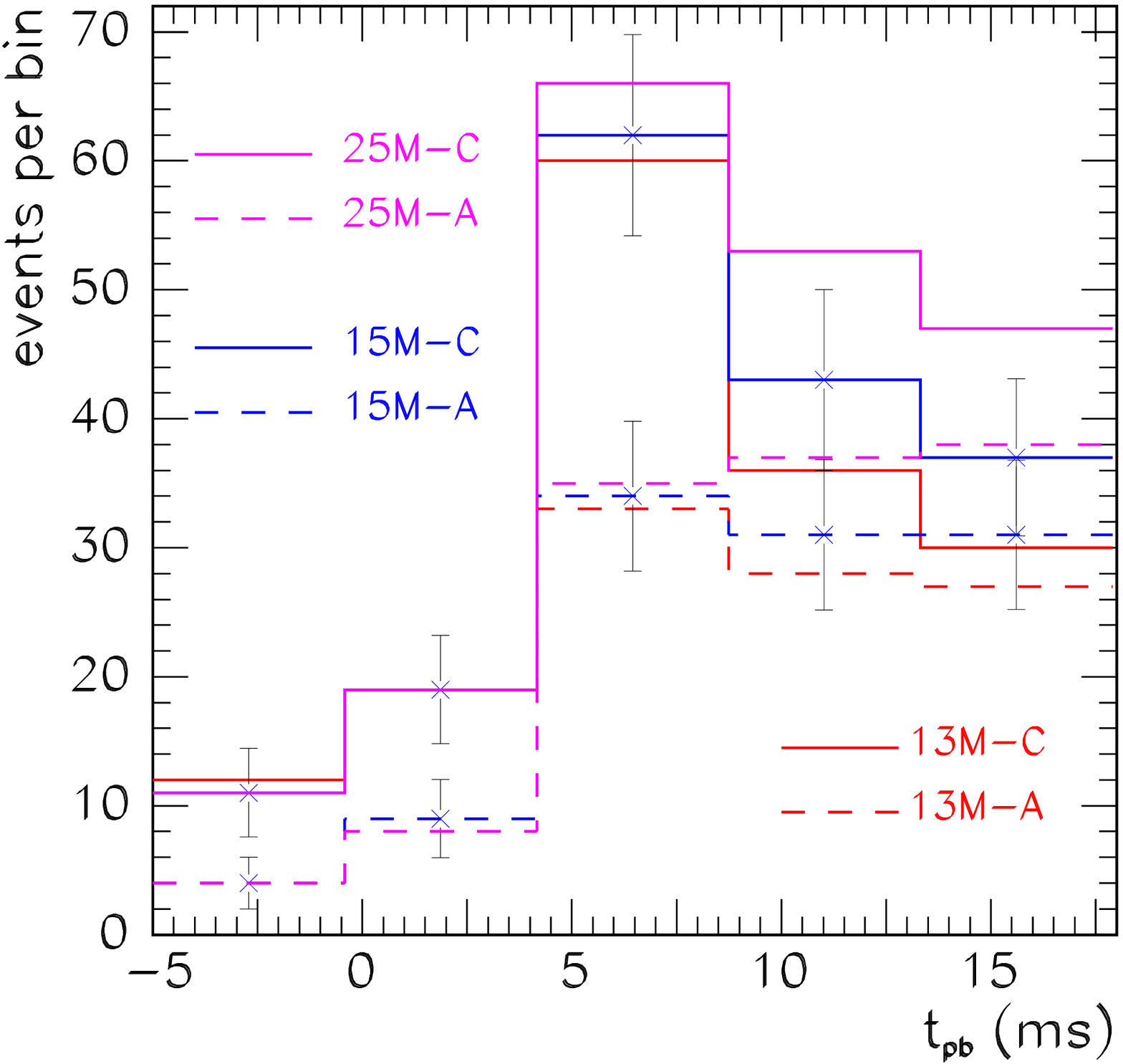,width=8.cm,height=6.5cm}
\caption{
Left: Neutrino luminosities as function of time for different
progenitor masses. 
Right: Number of events per time bin for different reactions in a megaton
water Cherenkov detector for a SN at 10 kpc for cases A (dashed lines)
and C (solid lines) and for different progenitor masses. Statistical
errors are also shown for the $15~M_\odot$ case. From Ref.~[11].
\label{nuepeak}}
\end{figure}

The prompt neutronization burst takes place during the first $\sim$
25~ms after the core bounce with a typical full width half maximum of
5--7$\,$ms and a peak luminosity 
of 3.3--3.5$\times 10^{53}\,$erg$\,$s$^{-1}$. The striking
similarity of the neutrino emission characteristics 
despite the variability in the properties of the
pre-collapse cores is caused by a 
regulation mechanism between electron number fraction and
target abundances for electron capture which establishes similar
electron fractions in the inner core 
during collapse. This leads to a convergence of the structure
of the central part of the 
collapsing cores and only small differences in the evolution of
different progenitors until shock breakout.
The small dependence of the neutronization burst on, e.g., the progenitor
mass can be verified in Fig.~\ref{nuepeak}, left panel (cf. also
Refs.~\cite{ourpaper,Takahashi:2003rn}). 

Theoretically, the identification of the neutronization burst is
cleanest with a detector using the charged-current
absorption of $\nu_e$ neutrinos, the most abundant flavor during the
burst. Examples of such detectors are 
heavy water detectors like SNO or liquid argon detectors like
ICARUS~\cite{Gil-Botella:2003sz}.
The simplest possible observable to identify the neutronization burst
is the total number of $\nu_e$ events within an arbitrary fixed period
$t_{\max}$ after the on-set of the neutrino signal.
However, the probability that the SN is obscured by dust is as high as
$\sim$75\%. Without an estimate for the SN distance, 
the total number of events observed cannot be connected to the SN
luminosity and is thus not a useful observable.   
Instead, the time structure of the detected neutrino signal should be
used as signature for the neutronization burst~\cite{ourpaper}. 

Since the event number in current and proposed charged-current
detectors is not high enough to allow for a detailed time analysis, 
we discuss only the case of a megaton water Cherenkov detector.
A draw-back of this choice is that this detector type does 
not have a clean signature for the $\nu_e$ channel. 
Instead, one has to consider the $\nu_e$ elastic scattering on
electrons, which is affected by
several backgrounds like inverse beta decay or reactions on
oxygen. In Ref.~\cite{ourpaper}, it was shown that this background can
be substantially reduced by using angular and energy cuts, as well as
Gadolinium to tag neutrons from inverse beta decays. 
The sample of elastic scattering events still contains the irreducible
background of scattering on electrons of other neutrino flavors than
$\nu_e$, but this contamination does not affect the possibility to
disentangle the different neutrino scenarios.

The time evolution of the signal depends strongly on the neutrino
mixing scheme. In case A, the $\nu_e$ survival probability  is
close to zero, and therefore the peak structure observed in the initial
$\nu_e$ luminosity is absent. On the contrary, in case C, 30\% of the
original $\nu_e$ remain as $\nu_e$  whereas 70\%
are converted into $\nu_x$. Since the cross section of $\nu_e$ on
electrons is much larger than that of $\nu_x$, the signal is dominated
by the contribution of $\nu_e$. These $\nu_e$'s follow the time evolution of
$L_{\nu_e}$, and thus lead to a clear peak in the signal.

In the right panel of Fig.~\ref{nuepeak} we show the expected neutrino
signal from $t=-5$ to 18~ms for different progenitor masses, and for
the mixing scenarios A and C. The peak structure can be clearly seen
in case C, but not in case A~\cite{ourpaper}.  Including
recent improvements of the electron capture rates or uncertainties in
the nuclear equation of state has only little effect on the
neutronization peak  compared to the size of the
statistical fluctuations. 
Therefore the observation of a peak in the
first milliseconds of the neutrino signal would rule out the normal mass
hierarchy with ``large'' $\theta_{13}$ (case A), breaking the 
degeneracy between scenario A and C observed in the $\bar\nu_e$ channel.

After the neutrino mixing scheme has been established, the robustness
of the theoretically predicted event number of the neutronization burst
makes a measurement of the distance to a SN located at 10
kpc feasible with a precision of about 5\%. Since it is likely that a
Galactic SN is optically obscured by interstellar dust and no other
distance determination can be used, this new method relying only on
neutrinos is very promising.

\section{Summary}

A reliable determination of neutrino parameters using SN neutrinos
should be independent from the primary neutrino fluxes produced during
the accretion and cooling phase of the SN. 
Earth-matter effects and the passage of SN shocks through the
H-resonance both introduce unique modulations in the neutrino energy
spectrum that allow one their identification without knowledge of the
primary neutrino spectra.
While the observation of Earth-matter effects in the $\bar\nu_e$
energy spectrum rules out case B, modulations in the $\bar\nu_e$ time
spectrum identify  case B. If the value of $\theta_{13}$ would be
known to be large, then the neutrino mass hierarchy would be
identified. Otherwise, the detection of the neutronization $\nu_e$
peak---a robust feature of all modern SN simulations---can break the
remaining degeneracy between A and C.

%%%%%%%%%%%%%%%%%%%%%%%%%%%%%%%%%%%%%%%%%%%%%%%%%%%%%%%%%%%%%%%%%%%%%%%%
\section*{Acknowledgments}
%%%%%%%%%%%%%%%%%%%%%%%%%%%%%%%%%%%%%%%%%%%%%%%%%%%%%%%%%%%%%%%%%%%%%%%%

We would like to thank our co-authors R.~Buras, A.~S.~Dighe,
H.-Th.~Janka, A.~Marek, G.~G.~Raffelt, M.~Rampp and L.~Scheck  
for many useful discussions and pleasant collaborations,
and T.~Schwetz for helpful comments about the manuscript.
MK acknowledges support by an
Emmy-Noether grant of the Deutsche Forschungsgemeinschaft and RT by a
Marie-Curie-Fellowship of the European Community.

\end{document}